\documentclass[conference]{IEEEtran}
\IEEEoverridecommandlockouts
\usepackage[utf8]{inputenc}
\usepackage{amsmath, amssymb}
\usepackage{graphicx}
\usepackage{float}
\usepackage{cite}
\usepackage{placeins}
\usepackage{caption}
\usepackage{subcaption}
\usepackage{url}
\usepackage{hyperref}
\usepackage{xcolor}
\def\BibTeX{{\rm B\kern-.05em{\sc i\kern-.025em b}\kern-.08em
    T\kern-.1667em\lower.7ex\hbox{E}\kern-.125emX}}

\begin{document}

\title{Flowing Through Hilbert Space: Quantum-Enhanced Generative Models for Lattice Field Theory\\
\thanks{This manuscript has been authored by UT-Battelle, LLC, under Contract No. DE-AC0500OR22725 with the U.S. Department of Energy. The United States Government retains and the publisher, by accepting the article for publication, acknowledges that the United States Government retains a non-exclusive, paid-up, irrevocable, world-wide license to publish or reproduce the published form of this manuscript, or allow others to do so, for the United States Government purposes. The Department of Energy will provide public access to these results of federally sponsored research in accordance with the DOE Public Access Plan.}
}

\author{\IEEEauthorblockN{1\textsuperscript{st} Jehu Martinez}
\IEEEauthorblockA{\textit{Department of Physics and Astronomy} \\
\textit{Texas A\&M University}\\
College Station, USA \\
jehumartinez12@tamu.edu}
\and
\IEEEauthorblockN{2\textsuperscript{nd} Andrea Delgado}
\IEEEauthorblockA{\textit{Physics Division} \\
\textit{Oak Ridge National Laboratory}\\
Oak Ridge, USA \\
delgadoa@ornl.gov}}


\maketitle

\begin{abstract}
Sampling from high-dimensional and structured probability distributions is a fundamental challenge in computational physics, particularly in the context of lattice field theory (LFT), where generating field configurations efficiently is critical, yet computationally intensive. In this work, we apply a previously developed hybrid quantum-classical normalizing flow model to explore quantum-enhanced sampling in such regimes. Our approach embeds parameterized quantum circuits within a classical normalizing flow architecture, leveraging amplitude encoding and quantum entanglement to enhance expressivity in the generative process. The quantum circuit serves as a trainable transformation within the flow, while classical networks provide adaptive coupling and compensate for quantum hardware imperfections. This design enables efficient density estimation and sample generation, potentially reducing the resources required compared to purely classical methods. While LFT provides a representative and physically meaningful application for benchmarking, our focus is on improving the sampling efficiency of generative models through quantum components. This work contributes toward the development of quantum-enhanced generative modeling frameworks that address the sampling bottlenecks encountered in physics and beyond.
\end{abstract}


\section{INTRODUCTION}

Lattice Field Theory (LFT) provides a powerful framework for studying non-perturbative quantum field theories, particularly in high-energy physics. It is widely used to model complex interactions in Quantum Chromodynamics (QCD), where the behavior of quarks and gluons is governed by interactions mediated by the strong force. A significant computational challenge in LFT is the efficient simulation of field configurations, which traditionally relies on Markov Chain Monte Carlo (MCMC) methods. While MCMC is effective, it suffers from critical slowing down, leading to large computational costs as lattice size increases~\cite{b1}. As a result, there is a pressing need for more efficient sampling techniques in LFT simulations. Recent advances in machine learning (ML) have introduced new possibilities for accelerating simulations. Specifically, normalizing flows (NFs)~\cite{b2,b3,b4,b5}, a class of generative models based on invertible neural networks, have shown promise in learning complex probability distributions and enabling more efficient sampling. Classical NFs, however, often require a large number of layers to capture the complexity of LFT field configurations, making them computationally expensive~\cite{b6,b7} and resource-intensive. 

To address these challenges, we explore the use of hybrid quantum-classical models, specifically the \textbf{Hybrid Quantum-Classical Normalizing Flow (HQCNF)}~\cite{b8, b9}. The hybrid model integrates quantum transformations, implemented via parameterized quantum circuits, with classical neural networks. This approach leverages the power of quantum entanglement and amplitude encoding to enhance the expressivity of the generative model, while classical layers provide adaptive coupling and potentially mitigate hardware imperfections. By integrating quantum components into the flow architecture, the HQCNF model reduces the number of classical layers, potentially lowering the computational cost while maintaining or improving performance.

In this work, we focus on applying the HQCNF model to LFT, specifically generating field configurations from scalar field theories. Our goal is to investigate whether quantum-enhanced NFs can improve the efficiency of field configuration generation by leveraging quantum components to capture non-classical correlations that are difficult for classical models. We demonstrate that by embedding quantum circuits into the flow architecture, HQCNF provides a more compact and expressive generative model, reducing the classical complexity required in traditional methods.

This paper presents an implementation of HQCNF for simulating scalar $\phi^{4}$ field theory on a two-dimensional $8\times 8$ lattice. We compare the performance of HQCNF against a classical NF baseline and evaluate the model using a range of diagnostics: effective action analysis, field value distributions, and two-point correlation functions. Our results suggest that HQCNF can achieve comparable or improved accuracy relative to classical models, while reducing the number of required layers, thus offering a potential path to more efficient simulation methods in quantum field theory.

By combining the expressivity of quantum circuits with the traceability of classical NFs, we highlight the potential of hybrid quantum-classical models for addressing computational bottlenecks in physics simulations. This work contributes to the ongoing exploration of quantum-enhanced generative models and sets the stage for their application to more complex quantum field theories and larger lattice sizes.

\section{BACKGROUND}
\label{sec:background}
Quantum machine learning has emerged as a promising avenue for enhancing computational techniques in high energy physics~\cite{b10}, particularly for generative modeling tasks. Generative models aim to approximate complex probability distributions, which is a fundamental challenge in various fields, including LFT. LFT provides a non-perturbative approach to quantum field theory, essential for modeling interactions in QCD, but requires efficient sampling methods due to the high-dimensional and correlated nature of the field configurations. Classical methods, such as MCMC, are often employed to sample these distributions. However, MCMC suffers from critical slowing down, especially in large systems, leading to prohibitive computational costs and limiting the practical application of LFT to larger lattice sizes or more complex theories.

Machine learning techniques, particularly NFs~\cite{b2,b3,b4,b5}, have gained attention as alternatives to traditional sampling methods. NFs are invertible transformations that map a simple distribution (e.g., Gaussian noise) to a complex target distribution through a series of bijective functions. The ability of NFs to learn complex distributions has made them a strong candidate for tasks such as density estimation and generative sampling. However, despite their promise, classical NFs often require a large number of coupling layers to model complex target distributions accurately. This results in high computational costs, especially for high-dimensional fields, making classical NFs less efficient for problems like LFT, where the target distribution exhibits strong correlations.

Recent advances in quantum computing have introduced the potential for quantum-enhanced generative models, which leverage the principles of quantum mechanics to enhance the expressivity and efficiency of classical machine learning models. Quantum circuit Born machines (QCBMs)~\cite{b11, b12}, quantum generative adversarial networks (QGANs)~\cite{b13,b14, b15}, and hybrid quantum-classical normalizing flows (HQCNFs)~\cite{b9} are examples of quantum-enhanced models that integrate quantum circuits within classical learning frameworks. These models utilize quantum circuits as nonlinear transformations or stochastic modules to enhance classical learning tasks by introducing non-classical correlations that classical systems cannot efficiently model. 

In particular, HQCNFs combine the flexibility of classical NFs with the representational power of parameterized quantum circuits. These hybrid models aim to reduce the computational complexity of generating field configurations by offloading part of the transformation to quantum circuits, which can capture entanglement and superposition in ways that classical models cannot. The quantum component of HQCNFs typically consists of parameterized variational circuits, which provide a trainable quantum transformation. Classical layers are responsible for learning the coupling between the transformed latent variables and adjusting the final output to match the target distribution.

The idea of combining quantum circuits with classical NFs is not new; previous work has focused on using quantum components for image generation tasks, demonstrating quantum-enhanced performance in scenarios such as low-resolution image reconstruction. However, extending this framework to LFT applications presents a unique challenge, as LFT requires models that can capture not only the marginal distributions of individual lattice sites but also higher-order correlations and non-local interactions that are characteristic of quantum field theories. 

While quantum-enhanced generative models hold potential, their practical application to LFT simulations is still in its infancy. Recent studies have demonstrated the use of quantum circuits to enhance classical neural networks for various tasks, but much of the work has focused on simpler systems or small lattice sizes. The key challenge for LFT is to scale these models to larger lattices and more complex field theories, where classical methods still face significant limitations. This work aims to address these challenges by testing a hybrid quantum-classical model on a simple $\phi^{4}$ field theory, with the goal of demonstrating the feasibility of quantum-enhanced sampling in high-energy physics simulations.

By exploring the use of HQCNFs in LFT, this paper seeks to extend the application of quantum machine learning models to a physically motivated domain, leveraging the strengths of both quantum and classical components to overcome the sampling bottlenecks inherent in traditional methods. Our approach builds on previous work in quantum-enhanced normalizing flows~\cite{b9}, where quantum circuits are used to enhance the learning capacity of generative models, but we focus specifically on the application of these models to the field configuration generation problem in LFT.

\section{METHODS}
\label{sec:methods}

In this section, we describe the formulation and implementation of the HQCNF model, which combines quantum circuits with classical invertible transformations to generate field configurations in LFT. The goal is to efficiently approximate the target distribution $p(\phi)\propto exp(-S[\phi])$, where $s[\phi]$ is the Euclidean action of the field configurations. The HQCNF framework consists of alternating classical and quantum transformations, leveraging quantum entanglement and classical affine coupling layers to enhance the expressivity and efficiency of the model.

\subsection{Model Formulation}

The objective of our approach is to learn a generative distribution $q_\theta(\phi)$ that approximates a target distribution $p(\phi)$. We construct $q_\theta(\phi)$ using a hybrid normalizing flow model $f_\theta$, that maps a simple base distribution $z \sim \mathcal{N}(0, \mathbb{I})$ to the complex target distribution over the lattice field configurations $\phi = f_\theta(z)$. The log-likelihood of a generated sample is given by the change-of-variables formula:
\begin{equation}
\log q_\theta(\phi) = \log \mathcal{N}(z) - \log \left| \det \left( \frac{\partial f_\theta^{-1}(\phi)}{\partial \phi} \right) \right|.
\end{equation}

\begin{figure}[h]
    \centering
\includegraphics[width=0.51\textwidth]{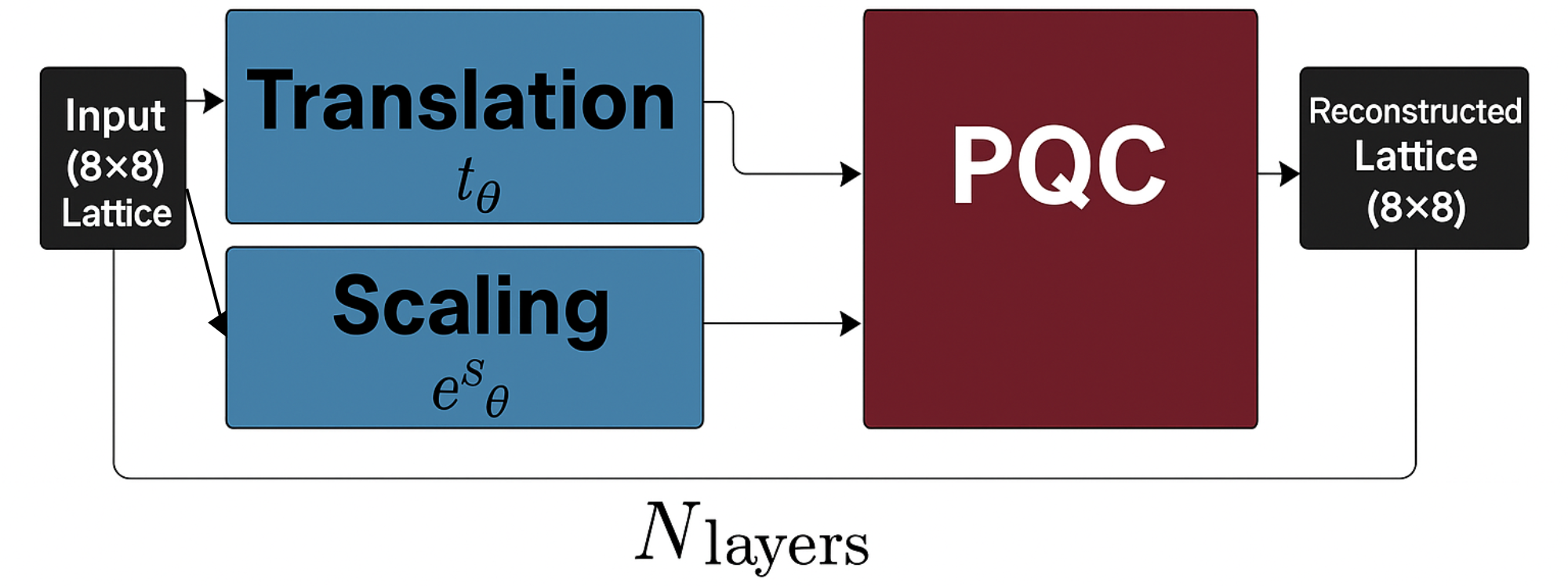}
    \caption{Schematic of the HQCNF model. The input $8 \times 8$ lattice field configuration is split into two components, which are processed by the classical part of the model using affine coupling layers with learned scaling and translation functions. The control parameters used in these classical transformations are then passed to the quantum component, which applies amplitude embedding and parameterized quantum rotations. The process is repeated over $N_{layers}$ number of layers (2 in this work), enabling the model to iteratively learn the correlations present in the target field theory.}
    \label{fig:hybrid_nf_diagram}
\end{figure}

Here, $f_\theta$ is a sequence of $K$ invertible transformations, each consisting of either classical affine coupling layers or quantum-enhanced transformations (Fig.~\ref{fig:hybrid_nf_diagram}). The series of transformations $f_{\theta}$ are applied to the latent variables $z$ to produce the field configurations $\phi = f_{\theta}(z)$, where each transformation step captures a more complex feature of the distribution:

\begin{equation}
f_\theta = f_K \circ f_{K-1} \circ \dots \circ f_1.
\end{equation}

\subsection{Classical Coupling Layers}

Each classical affine coupling layer splits the input $x \in \mathbb{R}^d$ into two partitions $x = (x_a, x_b)$ and applies an invertible transformation to the second part:

\begin{align}
y_a &= x_a, \\
y_b &= x_b \odot \exp\left(s_\theta(x_a)\right) + t_\theta(x_a),
\end{align}

where $\odot$ represents element-wise multiplication, and $s_\theta(x_a)$ and $t_\theta(x_a)$ are scale and translation functions parameterized by neural networks. The functions $s_\theta(x_a)$ and $t_\theta(x_a)$ are learned during training, allowing the model to adapt the transformations to the target distribution. The Jacobian of this transformation is triangular, and its log-determinant is computed efficiently:
\begin{equation}
\log \left| \det \frac{\partial y}{\partial x} \right| = \sum_j s_\theta(x_a)_j.
\end{equation}

These affine coupling layers provide a flexible and computationally efficient way to transform the input distribution while preserving invertibility, essential for tractable density estimation.

\subsection{Quantum Transformation Layer}

In addition to the classical coupling layers, a quantum component is introduced to enhance the expressivity of the model. The quantum transformation operates on a subspace of the latent space $z_q \in \mathbb{R}^m$ using a parameterized quantum circuit. The input features are encoded into quantum states using amplitude embedding. A classical encoder $g_\theta(z_c)$ generates the gate parameters $\alpha$ for the quantum circuit, which is then applied to the latent vector $z_{q}$:

\begin{equation}
|\psi(z_q; \boldsymbol{\alpha})\rangle = U(\boldsymbol{\alpha}) |z_q\rangle.
\end{equation}

Here, $U(\alpha)$ is a unitary operation parameterized by $\alpha$, typically consisting of quantum gates such as \emph{RY} rotations and $CNOTs$. After applying the quantum circuit, the resulting state is measured, yielding a probability vector $p_{q}$ from which the quantum loss is derived. This quantum operation introduces non-classical correlations into the model, providing an expressive transformation of the latent variables that enhances the generative capacity of the flow.

\subsection{Training Objective}

\noindent The total loss function is given by
\begin{align}
\mathcal{L}(\theta) &= - \mathbb{E}_{z \sim \mathcal{N}(0, \mathbb{I})} \left[ \log q_\theta(f_\theta(z)) \right] \nonumber \\
&\quad + \lambda_q \mathcal{L}_q + \lambda_\text{var} (\sigma_q^2 - \sigma_p^2)^2 + \lambda_S (\bar{S}_q - \bar{S}_p)^2,
\end{align}
where $\mathcal{L}_q = -p_q[0]$ is the quantum loss derived from the measurement probability, $\sigma_q^2$ and $\sigma_p^2$ are the variances of generated and target samples, and $\bar{S}_q$, $\bar{S}_p$ are their average action values. The coefficients $\lambda_q$, $\lambda_\text{var}$, and $\lambda_S$ control the contribution of each term.

\begin{figure}[ht]
    \centering
    \begin{subfigure}[t]{0.48\columnwidth}
        \centering
        \includegraphics[width=\linewidth]{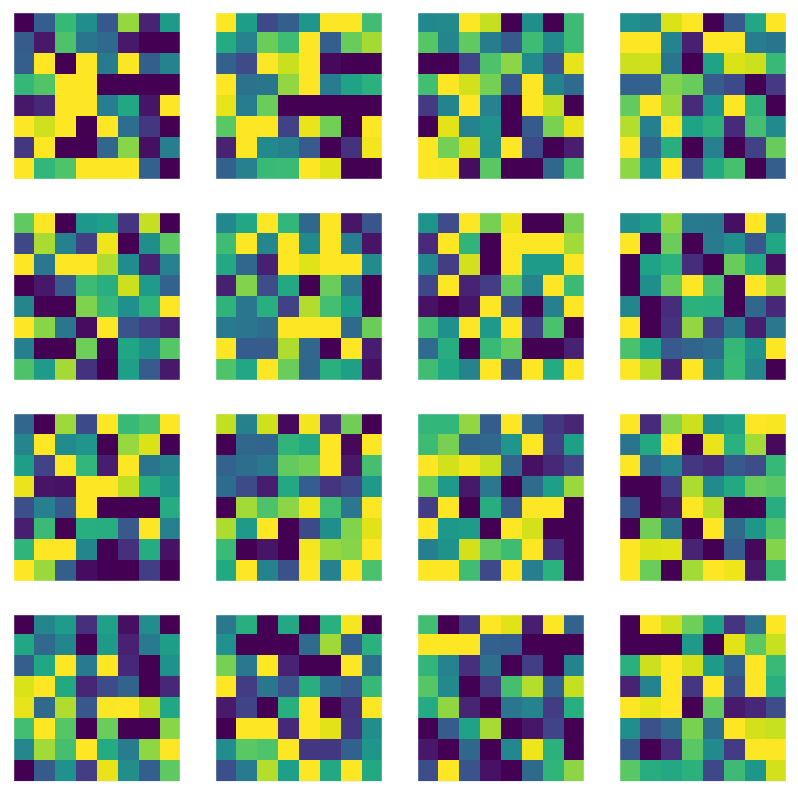}
        \caption*{(a) Samples from base distribution ($z \sim \mathcal{N}(0, \mathbb{I})$)}
    \end{subfigure}
    \hfill
    \begin{subfigure}[t]{0.48\columnwidth}
        \centering
        \includegraphics[width=\linewidth]{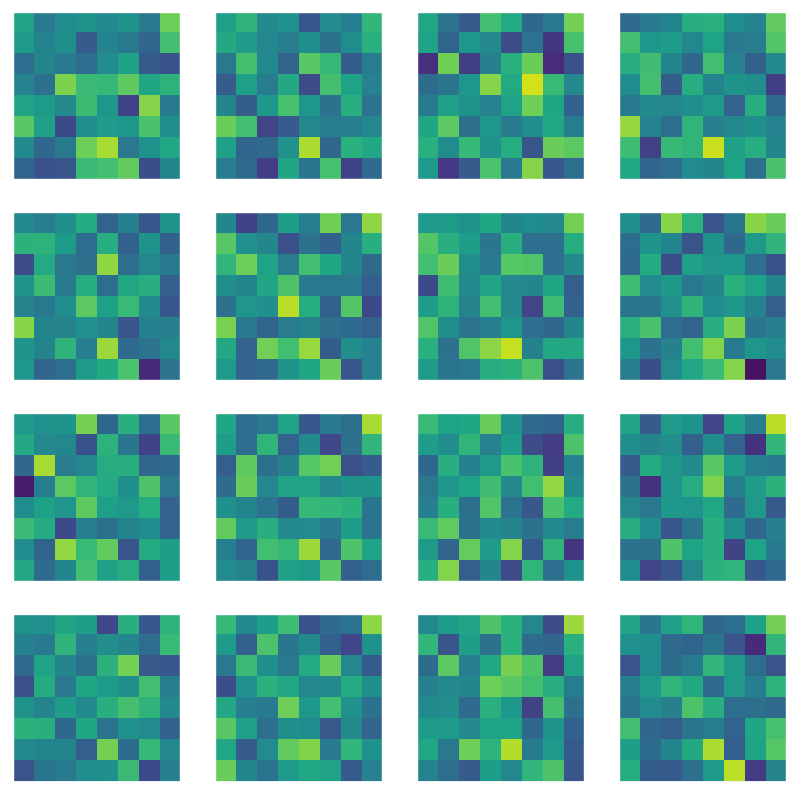}
        \caption*{(b) Samples from trained HQCNF model ($\phi = f_\theta(z)$)}
    \end{subfigure}
    \caption{
    Comparison between lattice field configurations sampled directly from the prior distribution (left) and those generated after training the hybrid quantum-classical normalizing flow model (right).
    The trained model produces configurations with smoother spatial structure and amplitude consistent with the target $\phi^4$ distribution.
    }
    \label{fig:config_comparison}
\end{figure}

It is important to emphasize that the HQCNF is not trained on image-label pairs, nor does it receive field configurations as input. Rather, the model samples latent vectors $z \in \mathbb{R}^{64}$ from a simple prior distribution $z \sim \mathcal{N}(0, \mathbb{I})$ (Fig.~\ref{fig:config_comparison} [left] ), and learns a transformation $f_\theta(z) = \phi$ such that the resulting outputs $\phi$ follow the statistics of a physical target distribution defined by a scalar field theory (Fig.~\ref{fig:config_comparison} [right]). The goal of training is therefore to align the distribution $q_\theta(\phi)$ induced by the model with the true distribution $p(\phi) \propto e^{-S[\phi]}$, based on statistical diagnostics rather than per-sample supervision.

To enable efficient computation of the Jacobian determinant, the flattened output field $\phi \in \mathbb{R}^{64}$ is partitioned into two components $(\phi_1, \phi_2)$, each of dimension 32. These parts are alternately transformed across the layers of the flow using scale-and-shift coupling functions, following the RealNVP architecture. This split allows the transformation to remain invertible and the Jacobian to remain triangular, enabling tractable and exact density evaluation during training (See Appendix in \cite{b9}).

\subsection{Evaluation Metrics}

To assess the model's ability to reproduce the target distribution, we evaluate:
\begin{itemize}
    \item The distribution of action values $S[\phi]$, as a proxy for the Boltzmann weight.
    \item Histograms of marginal field values $\phi(x)$ across all lattice sites.
    \item The two-point correlation function:
    \begin{equation}
    C(r) = \left\langle \phi(0) \phi(r) \right\rangle = \frac{1}{N} \sum_{i=1}^{N} \phi_i(0) \phi_i(r),
    \end{equation}
    averaged over samples and spatial directions.
\end{itemize}

These diagnostics provide a physically motivated measure of statistical similarity between the learned distribution $q_\theta(\phi)$ and the reference distribution $p(\phi)$ defined by MCMC.

\section{RESULTS}
    \label{sec:Results}

In this section, we evaluate the performance of the HQCNF model in generating field configurations for the scalar $\phi^4$ field theory on a two-dimensional $8 \times 8$ lattice. We compare the HQCNF model to a classical normalizing flow baseline, focusing on several key metrics: effective action, field value distributions, and two-point correlation functions. Additionally, we explore the computational efficiency of the HQCNF model by comparing it to the classical baseline in terms of the number of layers required for training and the quality of the generated field configurations.

The HQCNF model consists of two layers, alternating between classical affine coupling layers and quantum-enhanced transformations. Each affine layer contains a pair of neural networks implementing the scale and translation functions $s(\cdot)$ and $t(\cdot)$, each with three fully connected layers and $\tanh$ nonlinearities. These networks operate on partitioned inputs of size 32 and produce outputs of matching dimension. The quantum component is composed of a 5-qubit circuit using amplitude embedding, followed by two layers of parameterized $RY$ rotations and CNOT gates.

During training, latent vectors $z \sim \mathcal{N}(0, \mathbb{I})$ are passed through the flow to generate samples $\phi = f_\theta(z)$. The model is trained using the \emph{Adam} optimizer with a learning rate of $10^{-3}$ for 20 epochs. The quantum loss component is scaled by a multiplier of 100 to match the magnitude of classical loss contributions. The total number of trainable parameters is approximately  15,380, distributed across classical and quantum modules.

We compare our results against a classical normalizing flow baseline based on the model introduced in~\cite{b3}, which uses 16 affine coupling layers, each parameterized by a shallow feedforward neural network with hidden layers of width 8 and kernel size 3. The total number of parameters in the baseline is approximately 12,960. The model is trained using Adam with a fixed learning rate of 0.001 and batch size of 64 over 100 epochs per era for 25 eras, totaling 2,500 training epochs. 

\subsection{Effective Action Analysis}


\begin{figure}[ht]
    \centering
    \begin{subfigure}[t]{0.49\columnwidth}
        \centering
        \includegraphics[width=\textwidth]{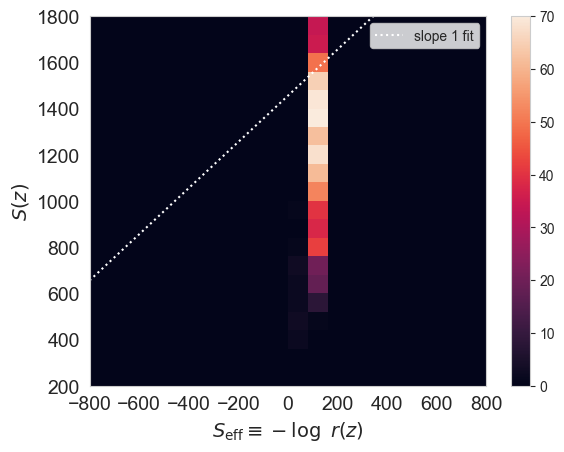}
    \end{subfigure}
    \hfill
    \begin{subfigure}[t]{0.49\columnwidth}
        \centering
        \includegraphics[width=\textwidth]{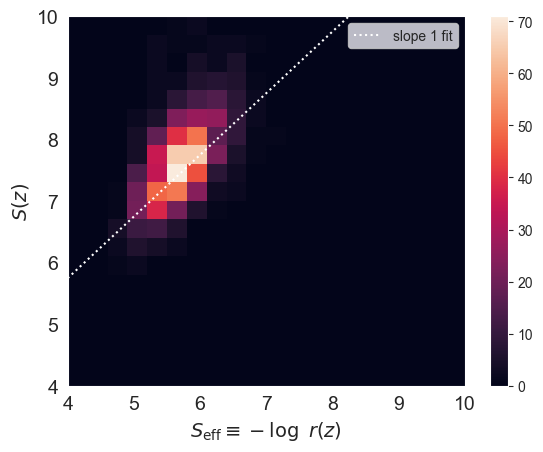}
    \end{subfigure}
    \caption{Correlation between the effective action $S_{\mathrm{eff}} = -\log q_\theta(\phi)$ and the true action $S(\phi)$ for generated samples.
    Perfect modeling would correspond to a slope-1 linear relationship, up to a constant shift.}
    \label{fig:action_correlation}
\end{figure}

To assess how well the HQCNF model captures the target distribution, we first examine the relationship between the \emph{effective action} of generated configurations,
\begin{equation}
S_{\mathrm{eff}}(\phi) \equiv -\log q_\theta(\phi),
\end{equation}
and compare it to the true scalar field action $S(\phi)$, given by

\begin{align}
S(\phi) &= \sum_{x} \left[ \frac{1}{2} \sum_{\mu} (\phi(x) - \phi(x+\hat{\mu}))^2 \right. \notag \\
&\left. + \frac{1}{2} M^2 \phi(x)^2 + \lambda \phi(x)^4 \right]
\end{align}

The effective action measures the log-probability of a sample under the model distribution $q_\theta(\phi)$ and serves as a surrogate for the model’s belief about the energy of the configuration. If the learned distribution closely approximates the true one, we expect a linear relationship of the form
\begin{equation}
S(\phi) \approx S_{\mathrm{eff}}(\phi) + \text{const.},
\end{equation}
indicating agreement up to a normalization constant.

Figure~\ref{fig:action_correlation} shows a 2D histogram of $S(\phi)$ versus $S_{\mathrm{eff}}(\phi)$. A strong correlation along the $y = x + b$ diagonal suggests that the model has learned an effective energy landscape consistent with the scalar $\phi^4$ theory. Deviations from this line indicate mismatches in variance or structural coverage.

\begin{figure}[h]
    \centering
    \includegraphics[width=0.49\textwidth]{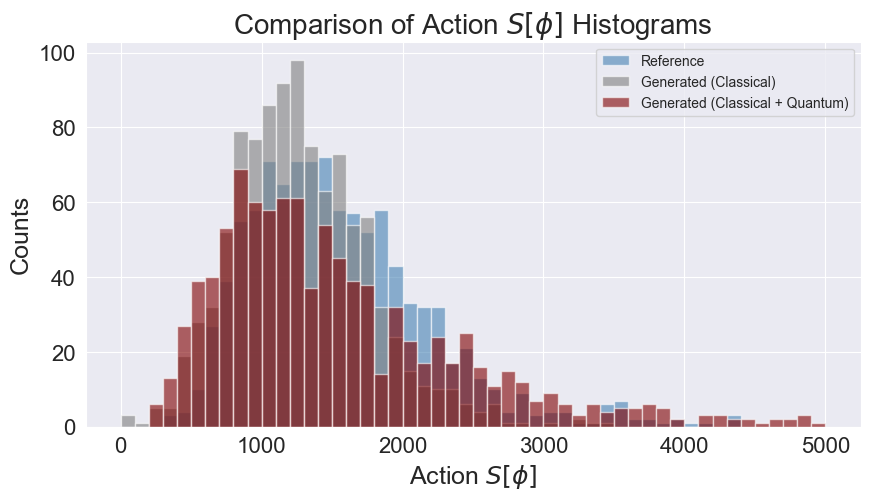}
    \caption{Histogram comparison of the action values $S[\phi]$ for reference and HQCNF-generated configurations.}
    \label{fig:action_histogram}
\end{figure}

To further evaluate the fidelity of the generated field configurations, we directly compare the distribution of action values $S[\phi]$ for both the HQCNF-generated and reference datasets. This provides a more global check of whether the model correctly samples from the Boltzmann-weighted distribution.
Figure~\ref{fig:action_histogram} shows histograms of the action $S[\phi]$ for both generated and reference configurations. A close match between these distributions suggests that the HQCNF has learned to reproduce the correct statistical structure of the target theory.

\subsection{Field Value Distribution}


Next, we examine the distribution of field values $\phi(x)$ across all lattice sites. The distribution of field values provides insight into the variance, symmetry, and heavy-tailed behavior of the field, which are critical characteristics of quantum field theories. In the target $\phi^{4}$ field theory, we expect a broad distribution with extended tails, characteristic of strong interactions between the scalar field.

\begin{figure}[h]
    \centering
    \includegraphics[width=0.45\textwidth]{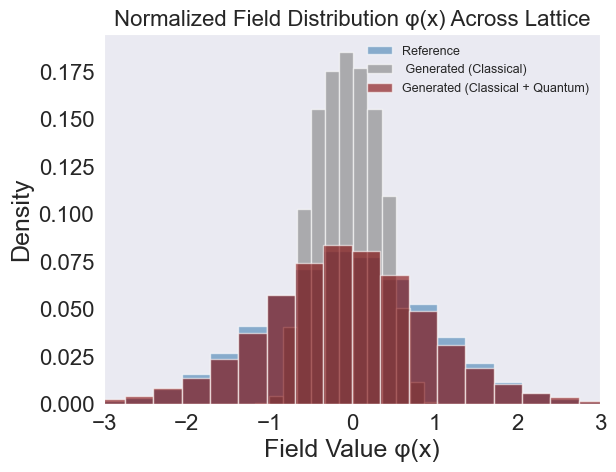}
    \caption{Distribution of scalar field values $\phi(x)$ across all lattice sites, comparing reference configurations and those generated by the HQCNF. The similarity between the two distributions suggests that the HQCNF captures the global statistical properties of the field.}

    \label{fig:field_hist}
\end{figure}

Figure~\ref{fig:field_hist} shows the normalized histogram of $\phi(x)$ for both the true (MCMC) and generated distributions. The reference distribution exhibits a broader shape with extended tails, characteristic of the strongly interacting $\phi^4$ field theory in the chosen parameter regime. The field values produced by the HQCNF follow this trend very closely. This is an improvement from the classical baseline model, that produced field values that are more sharply peaked around $\phi = 0$, with suppressed variance and a rapid falloff in the tails. This behavior indicates that while the model learns certain aspects of the spatial structure, it underestimates the amplitude of local fluctuations in the field. 

There are several possible causes for this discrepancy. First, mode collapse or over-regularization in the latent space can lead to low-variance outputs, particularly when the model learns a conservative approximation that avoids the higher-action tails of the true distribution. Second, generative models trained via maximum likelihood or reverse KL divergence tend to favor underdispersion, as assigning probability mass to rare or extreme events is penalized. Finally, the expressivity of the transformation function—especially in flows with fixed partitioning schemes—may limit the model's ability to reproduce heavy-tailed or multimodal features without deeper networks or auxiliary noise.


\subsection{Two-Point Correlation Functions}


The two-point correlation function is a central diagnostic in quantum field theory and statistical mechanics, as it quantifies the spatial structure and connectedness of fluctuations in the field. For a lattice scalar field $\phi(x)$, the two-point function is defined as
\begin{equation}
C(r) = \langle \phi(0)\phi(r) \rangle,
\end{equation}
where $r$ is the lattice separation and the average is taken over all configurations and equivalent lattice directions to improve statistics.

This quantity probes the decay of correlations with distance and is sensitive to key physical features such as mass gap, phase structure, and correlation length. For the scalar $\phi^4$ theory in the broken or near-critical regime, we expect nontrivial long-range correlations that deviate from purely exponential decay. Thus, reproducing the correct $C(r)$ behavior is essential for validating that the generative model has learned more than local or marginal statistics—it must capture nonlocal structure in the data distribution.

\begin{figure}[h]
    \centering
    \includegraphics[width=0.49\textwidth]{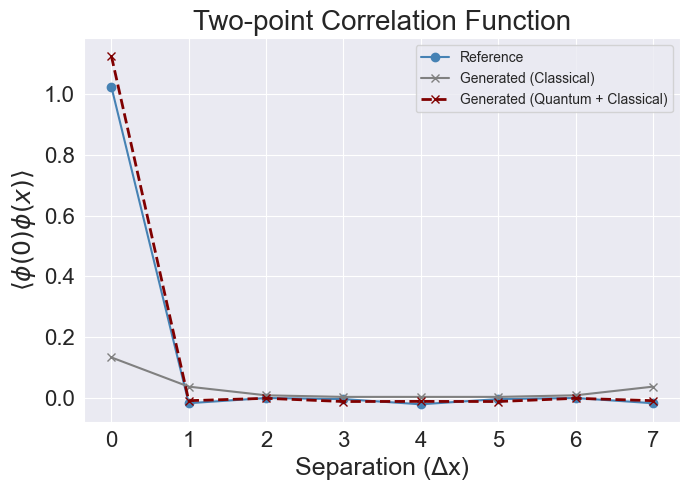}
    \caption{Two-point correlation function $\langle \phi(0)\phi(x) \rangle$ as a function of lattice separation $\Delta x$, comparing reference and HQCNF-generated configurations. The model captures short-range correlations and shows qualitative agreement with the long-range structure, reflecting its ability to reproduce nontrivial spatial dependencies.}
    \label{fig:2pt_corr}
\end{figure}

Figure~\ref{fig:2pt_corr} shows the two-point function computed on 500 configurations sampled from the HQCNF model, compared against configurations obtained via MCMC. We observe that the model accurately reproduces the shape and decay of correlations for all separations $r > 0$, suggesting that it successfully captures the local and medium-range structure of the field theory. This agreement is nontrivial and indicates that the flow has learned to encode physically consistent spatial dependencies.

However, a notable discrepancy arises at $r = 0$, where the HQCNF model overestimates the value of $C(0) = \langle \phi^2 \rangle$, the field variance. This behavior is more pronounced for the classical baseline model and is consistent with the narrower field distribution observed in the previous section. Since $C(0)$ contributes directly to the diagonal of the correlation matrix, a mismatch here reflects the generative model’s tendency to suppress field amplitudes.

Overall, the two-point function reveals that the HQCNF model captures spatial correlation structure remarkably well, while the classical baseline fails to reproduce the full scale of field fluctuations. This suggests that the model learns connectivity but not amplitude. 

\subsection{Moments of the Field}

We also compared scalar moments to evaluate how well the generated distributions align with known features of the theory. Table~\ref{tab:moments} presents the mean and standard deviation of $\langle \phi^2 \rangle$ and $\langle \phi^4 \rangle$.

\begin{table}[h]
    \centering
    \begin{tabular}{|c|c|c|c|}
        \hline
        \textbf{Moment} & \textbf{Reference} &\textbf{Generated (Classical)} & \textbf{Generated (Hybrid)} \\
        \hline
        $\langle \phi^2 \rangle$ & 1.0045  & 0.1360 & 1.1144 \\
        $\langle \phi^4 \rangle$ & 2.9829 & 0.0425 & 5.3992  \\
        \hline
    \end{tabular}
    \caption{Comparison of statistical moments of the scalar field between reference (true) and HQCNF-generated configurations. The values of $\langle \phi^2 \rangle$ and $\langle \phi^4 \rangle$ reflect key physical quantities of the field theory. While the HQCNF reproduces some qualitative behavior, discrepancies in the higher moments (especially $\langle \phi^4 \rangle$) highlight areas where the model can be improved.}
    \label{tab:moments}
\end{table}

These results provide further evidence that HQCNF is able to model relevant physical quantities effectively.

\subsection{Computational Efficiency}
Finally, we assess the computational efficiency of the HQCNF model relative to the classical baseline. The classical model requires 2500 epochs for training, each with a batch size of 64. In contrast, the HQCNF model achieves comparable results with only 20 epochs, demonstrating a significant reduction in training time.

Despite the reduction in the number of epochs, the HQCNF model produces field configurations with similar, or even improved, accuracy relative to the classical baseline. The quantum component allows the HQCNF model to reduce the number of required classical layers from 16 to just 2, significantly lowering the computational burden. This demonstrates the efficiency gains that can be achieved by leveraging quantum enhancements in the generative model. The fewer layers and shorter training duration make HQCNF an attractive alternative for large-scale simulations in high-energy physics.

% ...

% ...

\section{CONCLUSIONS}

The results presented in this work demonstrate the potential of the HQCNF model in generating lattice field configurations more efficiently than classical NFs. Our findings highlight several key advantages of the HQCNF approach. First, the HQCNF successfully learns the distribution of the $\phi^{4}$ field configurations and produces field distributions that closely match the reference (MCMC) distribution. This includes capturing the correct statistical properties of the field, such as variance and symmetry, while maintaining physical accuracy. The model accurately reproduces the shape and decay of correlations for all separations $r>0$, indicating that it captures both local and medium-range correlations. While there is a small overestimate in the correlation at $r=0$ (by approximately 10$\%$), the HQCNF model still captures the non-local dependencies inherent in the field theory, demonstrating its ability to model spatial dependencies accurately.

Another significant advantage of the HQCNF model is its computational efficiency. By leveraging quantum enhancements, we reduce the number of required classical layers from 16 to just 2, resulting in a substantial reduction in training time. The HQCNF model achieves similar or better accuracy compared to the classical baseline while reducing the number of epochs from 2500 to only 20, showcasing the potential for more efficient large-scale simulations in high-energy physics.

In terms of model complexity, the classical baseline for comparison uses convolutional neural networks (CNNs), which are commonly employed in classical normalizing flows for generating structured field configurations. In contrast, the HQCNF model uses fully connected neural networks (NNs), which simplify the model but also lead to a significant increase in the number of trainable parameters. We expect this architectural change to result in a large reduction in the number of parameters, making the model more scalable and efficient, particularly as we extend it to more complex field theories and larger lattice sizes.

The quantum-enhanced nature of the HQCNF model allows it to maintain expressivity while reducing the classical complexity involved in generating complex field distributions. This suggests that hybrid quantum-classical models are a promising path forward for addressing the sampling bottlenecks in simulations of quantum field theories.

While our results demonstrate strong potential, there are several avenues for further improvement. Scalability remains a challenge, as the current implementation is limited to two-dimensional scalar field theories on relatively small lattices. Extending the model to higher-dimensional systems or more complex gauge theories will require more expressive quantum circuits and potentially deeper neural networks. Additionally, the quantum circuit used in this study is simulated on noiseless backends, and implementing the HQCNF model on actual quantum hardware will require addressing hardware constraints such as noise, decoherence, and limited qubit count.

Future work will focus on addressing these scalability challenges, incorporating quantum error correction and noise mitigation strategies, and extending the model to more complex field theories, such as lattice QCD. We also plan to explore alternative quantum architectures, including variational circuits with higher entanglement and continuous-variable encodings, to further enhance the model's capability to handle larger systems and more complex interactions.

In conclusion, the HQCNF model represents a significant step toward quantum-enhanced sampling methods for high-energy physics. By combining the expressivity of quantum circuits with the tractability of classical neural networks, HQCNFs offer a flexible, efficient, and scalable solution for generative modeling in quantum field theory. As quantum hardware continues to evolve, hybrid models like HQCNF could become essential tools in the next generation of simulations, paving the way for breakthroughs in theoretical physics and beyond.



\section*{Acknowledgment}
\addcontentsline{toc}{section}{Acknowledgment}
This work was supported in part by the U.S. Department of Energy, Office of Science, Office of Workforce Development for Teachers and Scientists (WDTS) under the Science Undergraduate Laboratory Internships Program.(SULI). This work was partially supported by the U.S. Department of Energy, Office of Science, Office of Nuclear Physics Quantum Horizons: QIS Research and Innovation for Nuclear Science program at ORNL under FWP ERKBP91.

\end{document}